\documentclass{article}
\usepackage{amssymb}
\usepackage{graphicx}

\newcommand{\iec}{\mbox{i.\,e.\,}}

\newcommand{\egc}{\mbox{e.\,g.\,}}

\newcommand{\ket}[1]{\ensuremath{\left|  #1 \right\rangle}}

\newcommand{\op}[1]{\ensuremath{\widehat{\textsf{\ensuremath{#1}}}}}

\newcommand{\be}{\begin{equation}}
\newcommand{\ee}{\end{equation}}

\usepackage{chicago}

\begin{document}
\title{The sky is blue, and other reasons quantum mechanics is not underdetermined by evidence}
\author{David Wallace\thanks{Department of Philosophy/Department of History and Philosophy of Science, University of Pittsburgh; \texttt{david.wallace@pitt.edu}}}
\maketitle

\begin{abstract}
I criticize the widely-defended view that the quantum measurement problem is an example of underdetermination of theory by evidence: more specifically, the view that the unmodified, unitary quantum formalism (interpreted following Everett) is empirically indistinguishable from Bohmian Mechanics and from dynamical-collapse theories like the GRW or CSL theories. I argue that there as yet no empirically successful generalization of either theory to interacting quantum field theory and so the apparent underdetermination is broken by a very large class of quantum experiments that require field theory somewhere in their description. The class of quantum experiments reproducible by either is much smaller than is commonly recognized and excludes many of the most iconic successes of quantum mechanics, including the quantitative account of Rayleigh scattering that explains the color of the sky. I respond to various arguments to the contrary in the recent literature.
\end{abstract}

\section{Introduction: the underdetermination thesis}

\begin{quote}
I will do such things, --- \\
What they are, yet I know not: but they shall be\\
The terrors of the earth.

\begin{flushright}
W. Shakespeare, \emph{King Lear} (Act II, Scene IV)
\end{flushright}
\end{quote}

Feynman famously\footnote{In lecture 6 of the 1964 Messenger Lectures at Cornell University; see \texttt{https://www.feynmanlectures.caltech.edu/fml.html\#6}. The quote appears around 8:10.} quipped that nobody really understands quantum mechanics , but to judge from the recent philosophy literature the problem is not that we have \emph{no} way to understand it but that we have \emph{too many} ways. It has become common wisdom that we have multiple versions of quantum mechanics, and that these versions are not simply different interpretations of a common mathematical formalism, but different theories that have different formulations and make different ontological claims. Yet these theories (according to this common wisdom) are empirically equivalent: no experiment can tell them apart, at least in practice and perhaps in principle.

More precisely: even if we grant some standard assumptions of scientific realism --- scientific theories describe a mind-independent reality; scientific theories can be understood from a third-person view and are not irreducibly perspectival; science aims to describe the world and is not merely an instrumental calculus; primitives like `measurement' or `consciousness' have no place in the formulation of our theories --- it is widely claimed that we have multiple empirically-equivalent and mutually incompatible theories which reproduce the predictions of so-called `orthodox quantum mechanics'. The standard form of this claim is that there are three classes of scientifically-realist quantum-mechanical theories: Bohmian mechanics (aka the de Broglie-Bohm theory) as presented by, \egc, \cite{durrgoldsteinzanghibook,bricmontbook,durrlazarovickibook}; dynamical-collapse theories like the GRW theory \cite{grw}, the CSL theory~\cite{pearle}, and their generalizations and variants (reviewed in \cite{bassighirardireview}); the Everett interpretation in its modern decoherence-based form as presented by, \egc, \citeN{wallacebook} or Saunders~\citeyear{saundersroutledge1,saundersroutledge2}. Everett and Bohm are supposedly empirically indistinguishable in principle; the dynamical-collapse theories are normally said to be distinguishable in principle but not in practice. As such, quantum mechanics provides a real-life, non-contrived example of \emph{underdetermination of theory by evidence}, the failure of any realistically-obtainable evidence to distinguish between multiple scientifically-serious rival theories. 

Call this claim --- the tripartite underdetermination between Everett, Bohm, and dynamical collapse --- the \emph{quantum underdetermination thesis}, or QUT for short. It is very widely held: a non-exhaustive list of its defenders includes  \citeN{cordero2001}, \citeN{lewisontology}, \citeN{callenderquarantine}, \citeN{hoeferrealism}, \citeN{eggnospeculation}, and \citeN{eggsaatsi}, and slightly broader underdetermination claims including at least QUT are also defended by, \egc,  \citeN{barrettbesttheories}, \citeN{lyrequantum}, and \citeN{acunalandscape}.

If QUT is true, very substantial consequences follow for metaphysics, philosophy of science, and physics; I focus here on four. Firstly (as noted by all the authors I list above), QUT means that underdetermination of theory by evidence is not simply a philosopher's fantasy, as many philosophers of science (\egc \citeN{laudanleplinunderdetermination}, \citeN{nortonunderdetermination}) have claimed: it really occurs in science. As \citeN[p.45]{lewisontology} puts it, ``The three main versions of quantum mechanics \ldots are not mere hypothetical constructs, but genuine scientific theories, proposed, discussed, and compared in the physics journals ''. This reawakens the threat that underdetermination poses to the No Miracles Argument for scientific realism (that it would be a miracle for our best scientific theories to make such good predictions unless they were approximately true), and hence for scientific realism itself. If multiple incompatible theories all reproduce the myriad experimental predictions of quantum mechanics, all but one of them must be doing so despite being false; hence, miracles do happen after all; hence, the realist inference from spectacular empirical success to truth is illegitimate.

Secondly, if neither empirical evidence nor general considerations of realism help us to choose between versions of quantum mechanics, we are left with the so-called `extra-empirical virtues' often (e.g., Laudan and Leplin, \emph{ibid.}) appealed to by realists in response to the general threat from underdetermination. The choice between Everett, Bohm or Collapse will have to be made, if at all, on grounds like simplicity, elegance, parsimony, conformity  to the manifest image, locality, or fruitfulness in resolving other philosophical problems. The question of how to solve the measurement problem will become less of an interdisciplinary engagement with physics and more of a question in metaphysics: the question of how to understand the world through quantum mechanics can be expected to remain as contested, and as inconclusive, as traditional metaphysical debates about laws, properties, and the like. 

Thirdly, for that very reason, more empirically-minded philosophers of science will have reason  to recoil from the standard approaches to the measurement problem, committing us as they do to `speculation' \cite{eggnospeculation} or `deep metaphysics' \cite{saatsispin}, and to seek more nuanced or minimal understandings of QM that look to preserve what --- if anything --- is in common between them beyond the empirical data. Egg and Saatsi both make proposals along these lines, as do \citeN{cordero2001} and \citeN{hoeferrealism}, though see \cite{callenderquarantine} for scepticism that any such common core can be identified. (This third consequence is not cleanly separable from the first: if underdetermination threatens scientific realism, the threat can be defused by a more modest scientific realism that identifies a genuinely-explanatory common core of the various alternative theories and commits only to that core.)

These three consequences of QUT are quite widely recognized. A fourth consequence is less often remarked upon: if QUT is true, the physicist who insists that we should just `shut up and calculate', to the dismay of most philosophers of physics, gets the last laugh. The usual objection to that physicist's strategy is that physics is supposed to describe the world and not merely provide a calculational formalism; but if trying to do so leads us to underdetermination, if that underdetermination is irresolvable by the methods of physics and requires forever-inconclusive metaphysical speculation, if the very scientific-realist thesis that tells us that physics is supposed to describe the world is jeopardized  by QUT, and if erstwhile realists respond to that jeopardy by turning back to physics practice and trying to extract a more minimal core from it \ldots well, then, wasn't shutting up and calculating a good idea after all?

So: much depends on whether QUT is true. But QUT is not true. No extant version of Bohmian mechanics, and no extant version of dynamical collapse, can reproduce more than a tiny fraction of the empirical evidence that grounds QM. And so 
we do not after all have multiple theories which are empirically equivalent. And so any supposed consequences of there being such theories are moot.

I want to be as clear as possible about the claim here. \citeN{eggnospeculation}, in criticizing advocates of the view that there is no underdetermination at all, states (p.4) that those advocates `usually admit that there is some empirical equivalence between
the different versions of quantum mechanics', and then  immediately cites me as one of `the most outspoken advocates of this view'. I clearly was not outspoken enough, since I admit no such thing. I don't claim here that, as Egg (ibid., p.4) puts it, ``one particular version [of quantum mechanics is so much better than its alternatives that opting for it does not amount to speculation at all, but is simply what any scientifically reasonable person should do.'' I claim that Bohmian and dynamical-collapse theories flatly fail to reproduce all (or even most) of the data, and so just aren't available as empirically-adequate options in the first place, however scientifically reasonable or unreasonable we are.

The reason is simple to state. Quantum mechanics goes vastly beyond `non-relativistic quantum mechanics' (NRQM), the non-relativistic $N$-particle quantum theory for  which Bohmian mechanics and the best-known dynamical-collapse theories are formulated, and there has been at most extremely limited progress in extending either approach beyond NRQM; in particular, at least the great majority of predictions based on quantum field theory (QFT) are not reproducible within any known extension of Bohmian mechanics or within any known dynamical-collapse theory. So a huge fraction of the empirical predictions of quantum mechanics are reproduced by (at most) one of our three alternative theories; so there is no underdetermination. The purpose of this paper is to spell out this simple argument and to defend it against a variety of objections that have been made to earlier and less developed versions of the argument (\citeNP[pp.33-35]{wallacebook}; \citeNP{wallace-leedsrealism}).

The structure of the paper is as follows: in section \ref{what-is-required} I review exactly how Everett, Bohm and dynamical collapse solve the measurement problem, stressing the fact that arguably for all three, but certainly for the latter two, modelling of the measurement process (and not just the system being measured) is requried. In section \ref{adequacy-in-qft} I explain how this plays out in NRQM QFT, and argue that central features of NRQM that enable Bohmian and dynamical-collapse solutions to the measurement problem to work are absent in QFT. In section \ref{responses} I consider various responses to this argument that have been made (implicitly or explicitly) in the literature, finishing with the (curious) suggestion that reproducing the predictions of NRQM actually suffices for underdetermination and that getting the QFT predictions write is an extra-empirical virtue. In section \ref{scope-of-nrqm} I discuss so-called `non-relativistic' quantum phenomenology and demonstrate that a huge fraction of it actually requires QFT to be reproduced. In section \ref{research-programs} I consider the proposal that we think of Everett, Bohm and dynamical collapse as rival research programs, and argue that this does not give rise to a meaningful form of scientific underdetermination. In section \ref{conclusion}, the conclusion, I reflect on the actual significance of Bohmian mechanics and dynamical collapse in the foundations of quantum mechanics.

Three disclaimers before I begin. Firstly, I am interested specifically in the possibility of underdetermination between \emph{theories}; where I assume that different theories differ at least at the level of formalism. I am not concerned here with the possibility of multiple interpretations of a given formalism. This issue arises to some extent in considering various different interpretative moves in Bohmian and dynamical-collapse theories, but is particularly salient in the case of the Everett interpretation. The underlying formalism here is unitary quantum mechanics, understood as unsupplemented by hidden variables or a collapse rule, but this leaves open possibilities both for different metaphysics for the Everett interpretation (wavefunction realism vs spacetime-state realism, for instance) and for non-Everettian interpretations of unitary quantum mechanics (say, a single-universe understanding of consistent histories, or perhaps the relational account developed by Rovelli~\citeyear{rovelli-relational,rovellispaceisblue}). For the record I think that a properly structuralist metametaphysics should dissolve the first interpretative underdetermination and that the Everett interpretation is the unique coherent way to understand unitary quantum mechanics, but I won't argue for, or rely on, these points here. (See, respectively, \cite{wallacemathfirst} and \cite{wallacebook} for those arguments.)

Secondly, I confine my attention specifically to the supposed tripartite underdetermination between the Everett interpretation (or, more generally, unitary quantum mechanics), Bohmian mechanics, and dynamical-collapse theories, setting aside the possibility that, say, modal or retrocausal accounts of quantum mechanics can be constructed that deviate from the formalism of unitary quantum mechanics while remaining empirically equivalent to it. I do so for reasons of space and argumentative cohesion, and because that tripartite underdetermination is by far the most commonly discussed in the literature. That said, so far as I am aware there are no sufficiently-worked-out quantum-field-theoretic modal or retrocausal theories either, in which case most of the conclusions of this paper apply in those cases too.

Finally, I am assuming for the sake of argument that all the \emph{philosophical} problems in the three theories have been resolved. For instance, I am assuming that modern emergence-based accounts of macro-ontology, together with decoherence, solve the Everettian preferred-basis problem; that Bohmian mechanics and dynamical-collapse theories have some alternative account of macro-ontology that resolves the Everett-in-denial objection~\cite{deutschlockwood,brownwallace} and defuses the problem of tails~\cite{shimonydesideratum,albertloewer1996,corderogrw,mcqueenfourtails}; and that there are satisfactory understandings of probability in all three theories. I do so partly to conform to the norm in literature discussions of quantum underdetermination but mostly because failure by one or more of the theories to solve the measurement problem can only decrease the level of underdetermination. If only one of our three theories is empirically adequate, and we have to discard that theory because it fails to solve the measurement problem, the other two theories do not miraculously become empirically adequate. One minus one is zero, not two.

\section{What does empirical adequacy require?}\label{what-is-required}

The empirical predictions of quantum theory all follow, directly or indirectly, from the Born probability rule, which tells us both the possible outcomes of a measurement of any physical quantity (\iec, the eigenvalues of the associated self-adjoint operator) and the probabilities of each outcome obtaining (\iec, the summed mod-squared amplitudes of each eigenvalue's associated eigenvector(s) in an expansion of the quantum state). The Born rule can be applied to whatever physical quantity we like, and indeed in physical practice is routinely applied to many: scattering experiments mostly measure momentum; spectral line measurements measure energy; Stern-Gerlach experiments measure spin; two-slit experiments measure position (to say nothing of the various correlation functions calculated via the Born rule and used to deduce thermodynamic and transport properties in condensed-matter physics).

Naively we might interpret these measurements as, well, measurements: accurate determination of already-possessed properties. But of course quantum mechanics cannot  straightforwardly be so understood, and attempts to do so --- by, say, assigning joint probabilities at a single time to values of non-commuting physical quantities, or at multiple times to a single physical quantity, leads to inconsistency. In unitary quantum mechanics we dodge the inconsistency by requiring the probabilities to be ascribed only when our measurement process makes an in-practice-indelible record of the outcome,  and recorded redundantly in vastly many degrees of freedom (in the microscopic state of the measurement device and/or in an external environment); in recent decades this longstanding practice has come to be described in the parlance of \emph{decoherence theory}\footnote{For reviews in the physics literature, see, \egc, \cite{zurek91}, (Joos~\emph{et al}~\citeyearNP{joosetal}), or \cite{schlosshauerbook}; in the philosophy literature, see \cite{bacciagaluppiencyclopedia} or \cite[ch.3]{wallacebook}.}. Attention to the physical process of measurement both shows why simultaneously measuring non-commuting quantities is impossible, and ensures that the inconsistencies that in principle arise in sequential measurements do not in fact arise. Measurement, in unitary quantum mechanics, is the transformation of a superposition of eigenstates of whatever quantity is measured into a superposition of decohered states, and then the reapplication of the Born rule to that second superposition. (The conceptual justification of that rule, at least within Everettian quantum mechanics, is of course the Everettian probability problem).

Bohmian mechanics proceeds differently: it discards the Born rule entirely for all quantities \emph{except position}. Faced with the apparent objection that we measure quantities other than position, Bohmians (\egc, \citeN[pp.48-9]{maudlinqmbook}) normally follow Bell's observation that 

\begin{quote}
in physics the only observations we must consider are position observations, if only the positions of instrument pointers \ldots …If you make axioms, rather than definitions and theorems, about the ‘measurement’ of anything else, then you commit 
redundancy and risk inconsistency~\cite[p.166]{bellimpossible}.
\end{quote}
The idea is that since any measurement must \emph{ultimately} be recorded in the positions of objects --- indeed, in the positions of macroscopic objects --- then (a) it will suffice to reproduce the empirical predictions if we can reproduce the predictions about positions, and precisely because that does suffice, (b) it is actively unwise to do more than that because in doing so we `risk inconsistency'. (In unitary QM, that risk is instead addressed by only applying the Born rule once decoherence has occurred.) 

 Many Bohmians go further and advocate understanding the theory --- and, ideally, any physical theory --- in terms of a `primitive' (Allori~\emph{et al}~\citeyearNP{goldsteincommonstructure}; \citeNP{alloriprimitive}; Esfeld~\emph{et al}~\citeyearNP{esfeldprimitive}) or `primary' \cite{maudlinney} ontology. In this approach to physical theories, for a physical theory to make contact with empirical data it must make predictions about the locations of macroscopic objects in space, and to do so it must understand those macroscopic objects as agglomerations of fundamental objects --- the primitive ontology of the theory. Applying this to Bohmian mechanics means taking the primitive ontology to be the particles, which are always determinately localized in space. It is assumed that we have fairly direct access to the locations of those particles at least when they are collected together in bulk --- that is, when the measurement result has been recorded at the macroscopic scale. The quantum state may be taken as representing laws, or properties, or additional unobservable entities, or to be metaphysically sui generis, but at any rate what we are measuring when we conduct a measurement in Bohmian mechanics is not any feature of the quantum state: it is the spatial distribution of the primitive ontology.

(It's worth pausing briefly to notice how instrumentalist this recipe can get. A process that mainstream physics would describe as the measurement of a system's energy, for instance, may in Bohmian mechanics be understood as `measuring' only the macroscopic location of the pointer on the dial which records that energy. Still, that is not a failure of empirical adequacy.)

The key observation here is that it is essential in Bohmian mechanics that we can model, not just the microscopic system being `measured', but the `measurement apparatus' itself, all the way up to its recording of the `measurement outcome' in macroscopic data. (Scare quotes because in each case Bohmians may question whether this really ought to be called a measurement). And this is accepted, even welcomed by Bohmians: as Maudlin puts it:
\begin{quote}
A precisely defined physical theory \ldots would never use terms like 
``observation,'' ``measurement,” ``system,'' or ``apparatus'' in its fundamental 
postulates. It would instead say precisely \emph{what exists} and \emph{how it behaves}. If this 
description is correct, then the theory will account for the outcomes of all 
experiments, since experiments contain existing things that behave somehow. 
Applying such a physical theory to a laboratory situation would never require one to 
divide the laboratory up into ``system'' and ``apparatus'' or to make a judgment about 
whether an interaction should count as a measurement. Rather, the theory would
postulate a physical description of the laboratory and use the dynamics to predict what  the
apparatus will (or might) do. Those predictions can then be compared to the data 
reported. [\cite[p.5]{maudlinqmbook}; emphasis in original.]
\end{quote}

Something fairly similar happens for dynamical collapse theories. There, there is no `probability rule' at all: no part of the formalism has an inherently probabilistic interpretation. Rather, the dynamics is stochastic, and is constructed so that on measurement, the quantum state of the measurement device ends up definite and the probability of a definite outcome matches the Born-rule probability calculated inside unitary quantum mechanics. This again requires that the measurement outcome be encoded in macroscopic data, because the dynamical-collapse rule is constructed so as only to trigger at macroscopic scales (or at any rate at scales large enough that quantum interference is not in any case empirically observable.

The common requirement for our three theories is this: even if the system to which we want to apply quantum mechanics is microscopic, then our application requires us --- at least schematically, at least in outline --- to model not only that system but the apparatus  which we use to measure the system and to magnify the outcome of the measurement to macroscopic scales, or to decoherent scales for unitary quantum mechanics. To be clear, I don't want to claim that this is some general requirement for \emph{any} physical theory: classical general relativity, for instance, gets by fine without being able to model the workings of the atomic clocks used to test time dilation or the interferometer arrays used to detect gravity waves, both of which require quantum mechanics to understand. Rather, it is forced on us by the specific nature of the quantum measurement problem and of the methods by which our three theories solve it.

To be precise, if $\op{O}$ is some physical quantity for a quantum system (taken, for simplicity, to be nondegenerate and discrete), with eigenvalues $\{o_i\}$ and corresponding eigenstates $\ket{o_i}$, we need to ensure that the dynamics of measurement can be modelled within the theory and has the form 
\be
\sum_i \lambda_i \ket{o_i} \otimes \ket{\mbox{ready}}\rightarrow \sum_i \lambda_i   \ket{\mbox{Measure }o_i}
\ee
where $\ket{\mbox{ready}}$ is the quantum state of the measurement apparatus before the measurement occurs and  $\ket{\mbox{Measure }o_i}$ are output states (of systen and measurement device, jointly) such that  $\ket{\mbox{Measure }o_i}$ and  $\ket{\mbox{Measure }o_j}$ are macroscopically distinguishable for $i\neq j$. If this is the case, all three of our theories solve the measurement problem in at least this case: macroscopically distinguishable states formally pick out widely separated regions of configuration space, which corresponds in unitary quantum mechanics to decoherent states (since the decoherence basis is basically coarse-grained positions), in Bohmian mechanics to a probability distribution over coarse-grained particle locations that corresponds to the original Born-rule probabilities (since Bohmian mechanics matches the Born-rule probability distribution over positions) and in dynamical-collapse theories to a collapse with probability $|\lambda_i|^2$ to macroscopically-definite state $\ket{\mbox{Measure }o_i}$ (since the collapse rule is constructed to bring about collapses of superpositions with respect to coarse-grained position.

All of this is, I hope, uncontentious --- even commonplace. But we will shortly see that its transfer from NRQM to QFT is anything but simple.

\section{Empirical adequacy in quantum field theory}\label{adequacy-in-qft}

Let's consider exactly what goes into making Bohmian mechanics, or (say) the GRW dynamical-collapse theory, able to solve the measurement problem in NRQM. Both theories break the democracy of Hilbert space and pick out one particular commuting family of dynamical quantities --- in each case, particle position. In Bohmian mechanics, we add hidden variables to the theory whose actual positions are what position measurements return, place a probability distribution over those positions that reproduces the Born rule at an instant, and add equivariant dynamics so that we continue to reproduce it at later (or earlier) times. In GRW, we introduce a collapse mechanism that suppresses superpositions with respect to position, and organize it so that its effect is significant only when the macroscopically-coarse-grained position of a collection of particles is in a superposition. (And other dynamical-collapse theories for NRQM work in broadly the same way.)

In this process, position plays two roles:
\begin{enumerate}
\item It is fundamental and exact. The whole point of a modificatory solution to the measurement problem, as advocated by everyone from Bell onward, is to replace imprecise, emergent, high-level, regime-dependent concepts in the formalism of a theory with sharply-stated axioms and dynamical laws. It would not, for instance, be acceptable to advocates of either theory to write a collapse law, or a hidden-variable theory, that directly references `the basis preferred by decoherence' --- decoherence is too high-level, too inexactly stated, too dependent on the particularities of the dynamical situation, to be referenced in the fundamental laws of a modified quantum theory. Position fits the bill admirably: it is written into the very foundations of NRQM, to the point that introductory presentations of NRQM (\egc \citeNP{Rae1992,griffithstextbook}) often start with the wavefunction-on-configuration-space way of formulating the theory and only later develop the general Hilbert-space framework.
\item It can be coarse-grained  to  obtain variables (like center-of-mass position) which suffice to individuate macroscopically distinct states (and of course the Born probability distribution over a coarse-graining of a set of commuting observables is just the coarse-graining of the Born probability distribution over the original observables). It is precisely because macroscopically distinct states have distinct coarse-grained positions that the Bohmian probability distribution and the dynamical-collapse law assign the correct (Born-rule) probabilities to each.
\end{enumerate}
The reason all this is possible --- and indeed, the reason primitive ontology is possible --- is that there is a relatively simple, direct relation between the \emph{fundamental} description of NRQM and the description of the \emph{macroscopic regime} in NRQM, with the latter simply describable in terms of coarse-grainings of properties of the former.\footnote{Or at least, let's stipulate that it is simply describable. I actually think that the relationship between the world of extended, colored, textured continua that describes our world at human scales is way more complicated than the simple agglomeration account presupposed by (some forms of) Bohmian and dynamical-collapse theories (on this point see also~\cite{battermanmiddleway}) but in accordance with my last disclaimer in section 1, I'll ignore this concern.} And the underlying reason why it is so hard to generalize Bohmian or dynamical-collapse theories to QFT is that in QFT, the relationshp between the microscopic theory and that theory's account of the macroscopic regime is dizzyingly indirect.

Let's see why.\footnote{The physics I discuss here is standard and I do not attempt to give original references. See \cite{wallaceqfthandbook}, and references therein, for details.} A relativistic quantum field theory, like quantum electrodynamics (QED) or the Standard Model, will have a `non-relativistic particle mechanics' regime. This regime requires (inter alia) that energies are low, radiation is negligible, and antimatter is absent; if these requirements are met, certain features of the QFT system can be described to a good degree of accuracy by NRQM. The precise details are beyond the scope of this article; what matters for our purposes is that the process is fairly well understood on a technical level.\footnote{A lightning-fast and highly incomplete account for those familiar with modern QFT: to get NRQM from QED, start with the path-integral and construct an effective field theory by carrying out the integral over the electromagnetic field explicitly (following, \egc, \citeN[ch.12]{breuerpetruccionebook}, then take $c\rightarrow \infty$. To get it from the Standard Model, first obtain QED-plus-nuclei as an effective field theory from the full Standard Model, then repeat the above procedure.}

Since the non-relativistic particle-mechanics regime is the regime in which the (slow-moving, macroscopically large) objects that comprise measurement results are described (or rather: since that is how relativistic QFT describes them), a hidden-variable or dynamical-collapse version of a relativistic QFT needs to be defined via a preferred set of dynamical quantities whose coarse-graining picks out the macroscopic NRQM states: non-relativistic particle position, most naturally, or perhaps something that reliably covaries (and quantum-mechanically commutes) with it. 

But the non-relativistic regime is not fundamental, even relative to a given QFT: it is an emergent, approximate, high-level, situation-dependent approximation to it. It is not the sort of thing that defines variables that can be used to define a fully general modification of the QFT, any more than the decoherence-preferred basis is. A microphysically satisfactory Bohmian or dynamical-collapse theory will need to find some microphysically-stateable, precisely-defined dynamical variable which, on coarse-graining and restriction to the non-relativistic particle-mechanics regime, nonetheless delivers coarse-grained particle position or some appropriate surrogate.

I do not know of any such dynamical variable, and a number of very general features of QFTs make me sceptical that any exists. It's easiest to illustrate this by looking at two commonly-discussed classes of Bohmian QFTs which have been proposed. The first class (e.g.\, D\"urr \emph{et al}~\citeyearNP{goldsteinqft03}) analyzes the QFT Hilbert space as a Fock space built from 1-particle states, and then chooses as its preferred basis the basis of definite-particle-number states, and within each definite-number subspace, the (Newton-Wigner) position basis. This gives a Bohmian QFT very close to standard Bohmian mechanics: $N$-particle states evolve under a relativistic version of the guidance equation, supplemented by a stochastic process that can create or destroy particles. This suggests in turn a relatively clear route from QFT to NRQM as a limiting case.

The first problem here is that \textbf{parameters in QFT are renormalized and scale-dependent}. The `bare parameters' --- the parameters that appear in a microphysically precise statement of a QFT --- are not directly measurable. They are `renormalized' by large dynamical effects happening at very high energies (the so-called `cutoff energy' of the QFT). The measured parameters are therefore in an important sense non-fundamental, being related to the fundamental parameters by complex dynamical processes. Furthermore, these processes are scale-dependent, so that the values of, say, the charge and mass of the electron vary according to the energies at which the electrons being studied interact. Since the mass and (in the presence of a magnetic potential) the charge appear in the Bohmian guidance equation, we have a choice between using the renormalized charge and mass at an appropriate scale (which seems to violate the requirement of microphysical precision) and using the bare charge and mass (in which case the theory will no longer be equivariant, since it is the renormalized charge and mass that appear in the non-relativistic Schr\"{o}dinger equation, and the `particles' being described will not be related in any simple way to the empirically-detected particles of NRQM).

To make matters worse, \textbf{particles in QFT are approximate and emergent}. In \emph{free} quantum field theories, there is a perfect duality between field and particle descriptions of the theory: we can construct the same QFT either by quantizing a classical linear field theory or by reinterpreting that field theory as a one-particle quantum theory and constructing the multiparticle Fock space for that theory. But in interacting QFT this breaks down, and the clear consensus in both physics and philosophy is that the particle description must be regarded as approximate. Partly this follows from general considerations about localizability, inequivalent representations, Haag's theorem, curved spacetime and the like (see \cite{fraserparticlesroutledge} and references therein) but more directly, it can be read off from physics practice. The variation of parameter energy I discuss above can be understood heuristically as occurring because particles polarize the quantum vacuum (or, still more heuristically, because they are accompanied by swarms of virtual particle-antiparticle pairs, though that is best read as a metaphor), and the division of an excitation into `particle' and `bits of polarization around a particle is somewhat arbitrary and, on any natural choice of convention, ends up being scale-dependent. So the `right' particle-theoretic description of a QFT is scale-dependent and hence a given analysis of QFT states in terms of particle states is an approximation applicable only at certain energies. 

In the most dramatic cases, different energy levels need to be analyzed not just with different particle parameters but with different particles altogether. In quantum chromodynamics the high-energy regime is naturally analyzed in terms of quarks and gluons; at lower energies this description breaks down entirely and needs to be replaced by a description in terms of protons, neutrons, and mesons. (It is true only in an approximate sense that these particles are composed of quarks; the proton, for instance, is more accurately describable as an excitation of a certain symmetrized triple product of the quark \emph{field}.)

So a Bohmian QFT based on particles fails to meet the core requirement that a physical theory should be sharply stateable in precise microphysical terms. An association of Bohmian particles to the `definite-particle-number' states of a QFT is an association of those particles that is approximate, fuzzily-stated, regime-dependent\ldots exactly the things that Bohmian mechanics is supposed to avoid.

A natural alternative to associating hidden variables to particles is to associate them to fields; if the fields are those associated with fermionic matter this would probably fail to reproduce the macroscopic regime, but the electromagnetic field does seem to be approximately classical in the non-relativistic regime. Bohmian QFTs based on this idea have been developed by \citeN{struyvewestman07}. However, \textbf{classical field states in QFT are also emergent}. The quantum version of a `classical' electromagnetic field is a coherent state, built in free-field theory as a superposition of definite-photon-number states. The free-field regime required for this analysis is isomorphic to a free-particle description and so coherent states are emergent and energy-scale-dependent in the same way that particles are. In particular, while in a free field theory it is fairly straightforward to write down a guidance equation for hidden variables assigned to field configurations such that the movement of the hidden variables guidance equation tracks the coherent states, I see no particular reason in an interacting field theory to suppose that such a hidden-variable assignment at the level of the `bare' (\iec, microscopic, fundamental) field configurations would bear the appropriate relation to the emergent, low-energy-scale, coherent-state description applicable to classical electromagnetic radiation.

These are not the only stratagies I am aware of for constructing Bohmian QFTs (see \cite{struyvereview} for a review of the subject, including several others based on fermionic fields) but the problems generalize: a Bohmian field theory has to select either a microscopically precise preferred basis (in which case there is no guarantee that it will correspond in the appropriate way to the basis of macroscopic non-relativistic coarse-grained positions required to make the theory empirically adequate) or else a basis directly characterised in terms of non-relativistic particle positions (in which case the theory fails to be microscopically precise). The problem is obscured by a focus on free-field theory or on a simplified treatment of interactions that does not properly engage with renormalization, but emerges in full force once we work in a realistic interacting quantum field theory. And the same is true \emph{mutatis mutandis} for dynamical-collapse theories, where the situation is if anything even worse: the state of the art is probably Tumulka's~~\citeyear{tumulka} relativistic version of GRW and the field-theoretic version of dynamical collapse developed by \citeN{bedinghampearle} (following earlier work by Bedingham~\citeyear{bedingham2010,bedingham2011}), both of which prioritize a particle description and neither of which even attempt to reproduce a realistic interacting QFT. 

But in fact the problem is even worse than this. I have written as if there is an unambiguous microscopic description of QFTs, but on the mainstream interpretation of QFT this is not so, because \emph{quantum field theories are effective theories}. There are indefinitely many ways to fill in the short-distance physics in, say, the Standard Model or QED, corresponding to the indefinitely many ways high-energy degrees of freedom can be truncated and the continuum infinity of scales at which that truncation occurs, and a central lesson of modern renormalization theory is that differences between these possibilities are invisible at the comparatively large scales at which we extract empirical information from these theories. A QFT is probably best understood as an equivalence class of theories reproducing the same large-scale particle physics phenomenology~\cite{wallaceconceptualqft}; an appropriate scientific realism for the effective-field-theory era is a scientific realism that takes seriously only that large-scale phenomenology and remains agnostic about short-distance physics~\cite{williamseffectiverealism}. But then it is unclear even how to state a satisfactory Bohmian or dynamical-collapse version of QFT: a microphysically-exact association of hidden variables or collapse rules would be associating them to a theory that is flatly fictitious. (I develop this concern in much more detail in \cite{wallace-leedsrealism}.) The extant literature on Bohmian or dynamical-collapse QFTs does not engage at all with these issues: indeed, it is pretty much concerned with QFT as it appeared in the 1930s, before even the formal developments of renormalization theory in the 1940s, let alone the revolution in our conceptual understanding brought by the renormalization group and the effective-field theory concept.

So I conclude that we have at present no version of QFT that realizes the goal of the Bohmian or dynamical-collapse programs and that is empirically adequate for QED, let alone the full Standard Model.  For the reader who disagrees, there is a simple way to respond: state a Bohmian or dynamical-collapse theory supposedly equivalent to QFT in microphysical terms, then analyze some QFT prediction --- the electron-positron scattering cross-section, say --- in that theory. Carry out the analysis to one-loop order, to make sure that renormalization issues are not being missed, and then track it --- in outline, perhaps, but in technical calculational detail and not just via a verbal gloss ----through  the measurement process up to the macroscopic record of the outcome. As we saw in section~\ref{what-is-required}, this is the criterion for empirical adequacy for a Bohmian or dynamical-collapse theory (or the Everett interpretation, arguably, but there the process is fairly clear) and it is satisfied in the non-relativistic regime. It needs to be satisfied in the QFT regime too if a proposed `alternative' to unitary QM does not consist merely of a recapitulation of unitary-QM calculations followed by a verbal assurance that \emph{of course} the proposed modifications of QM recover the correct macroscopic predictions.

No extant hidden-variable or dynamical-collapse theory fulfils this requirement for even one concrete QFT prediction (let alone all of them). So there is no underdetermination in quantum field theory.

(Before continuing, I want to note that the argument in this section makes \emph{no use at all} of the requirement that our physics ought to be Lorentz-covariant. This is one of the most commonly discussed worries about relativistic versions of hidden-variable or dynamical-collapse theories (see, \egc, \cite{myrvoldcollapse}, \cite{barrettbesttheories},  (D\"urr \emph{et al}~\citeyearNP{durretalrelativisticbohm}), \cite[ch.7]{maudlinqmbook}), but I am happy to disregard it. My concern is not that our empirically-adequate versions of Bohmian mechanics or GRW rely on a hidden notion of simultaneity or require action at a distance; it is that \emph{we have no such versions}.)

\section{Defending underdetermination}\label{responses}

I'm tempted to finish the paper here. The quantum underdetermination thesis, QUT, states that empirical predictions of QM can be reproduced equally well by dynamical-collapse theories, Bohm-type hidden-variable theories, or unitary quantum theory; unitary quantum theory makes a large number of predictions in the QFT domain that are not replicated by any extant dynamical-collapse or Bohmian theory; so QUT is false, end of story. But quite a number of authors have responses that defend QUT despite awareness of the QFT domain, and so there is a need to engage with their arguments; I do so in the remainder of the paper.

The first response is to flatly deny that we lack empirically-adequate alternatives to unitary QFT (normally the claim is made specifically for Bohmian mechanics; there seems to be a wider appreciation that QFT versions of dynamical-collapse theories are works in progress). \citeN[pp.310--11]{dieksrealism}, for instance, states that `a Bohmian version of quantum field [theory] can be developed'; \citeN[p.170]{bricmontbook} claims that `all the predictions of the usual quantum field theories are also obtained in those Bohmian-type models'. But these claims are not substantiated, and are instead backed by citations to the various QFT proposals we discussed in the previous section (whose authors are normally significantly more modest about what has been achieved). The acid test here is the one we just considered: if you want to claim that your preferred alternative to unitary QFT can obtain all of its predictions, \emph{actually do it} for at least a representative sample of them.

A more sophisticated concern is that the supposed empirical predictions of QFT are illusory, because QFT is in mathematically too bad shape to be treated as any sort of theory at all. \citeN[p.170]{bricmontbook} hints at this (`to the extent that those models are rather ill-defined mathematically, the same thing is
true for ordinary quantum field theories'); \citeN[p.193]{durrlazarovickibook} are explicit about it (`there does not exist a fundamental, mathematically coherent and consistent formulation of a relativistic quantum theory with interaction that could extend the
analysis of the foregoing chapters [of their book on Bohmian mechanics] to relativistic physics'). But these comments seem to be based on a very outdated picture of quantum field theory, in which renormalization is miraculous black magic, which has been obsolete for fifty years.  I'm not aware of any advocate of this response who engages critically with either the development of the effective field theory concept in physics in the 1970s or 1980s, or the developing consensus in more recent philosophy of physics that effective-field-theory approaches to QFT are legitimate and the old criticisms of  mainstream QFT are outdated (see, \egc, \cite{wallacecritique,williamsnaturalness,millerhaag,fraserinfrenchsaatsi,rivatgrinbaum} --- though see Fraser~\citeyear{Fraser2009,frasercritiqueresponse} and \citeN{kuhlmannaqft} for more skeptical views).  

(D\"urr and Lazavoricki provide an especially clear demonstration: they write (ibid., p.205) that 
\begin{quote}
the infinities which are so bothersome and which we would like to
sweep under the carpet will always resurface in one way or another --- at least they
have done up until now. At the end of the day we must realize that the problem is
not just a quest for the right mathematical language, but that relativistic quantum
theory will require new physical insights. The infinities which appear abundantly
in the programme are not merely mathematical problems that we can try to solve
by new techniques. They are clear signs of fundamental physical problems which
will lead to a distortion of the notion of physical theory if we keep trying to
push them to one side[.]
\end{quote}
The mainstream QFT community will \emph{agree with them} that the infinities of QFT require physical and not just mathematical insights, and will cite the development of the renormalization group and the effective-field-theory program as providing exactly those insights ---but D\"urr and Lazarovicki do not mention these developments, or indeed anything at all in QFT beyond the 1950s.)

A third alternative is to deny the coherence of the Everettian solution to the measurement problem. \citeN[p.7]{eggsaatsi} consider (though do not endorse) this move in their defense of QUT against the QFT objection: they observe that``one can question whether the Everett interpretation is even a candidate for a satisfactory view of the empirical world.'' Indeed one can; but what has this to do with underdetermination? Let's stipulate for the sake of argument that the Everett interpretation is utterly inadequate for physics and should be rejected wholesale.  Bohmian mechanics does not thereby acquire the ability to predict the magnetic moment of the electron. (Perhaps the idea is that once the Everett interpretation is rejected and it is acknowledged that we have to explicitly modify or supplement QM to solve the measurement problem, we will have underdetermination between the different modificatory or supplementation strategies? I suppose I concede that, but (a) this is not the normal form of QUT and (b) the empirical scope of any current strategy is pretty meagre; cf section \ref{scope-of-nrqm}.)

A final strategy is to acknowledge that it is an advantage of the Everett interpretation that it can recover the results of relativistic quantum field theory, but to argue that virtue trades off against other deficiencies. I confess to finding this so puzzling as a defense of \emph{the underdetermination thesis} that I will quote twol advocates of this strategy at length to try to avoid misunderstanding.

Consider first \citeN{callenderquarantine}, who introduces (p.60) the vivid metaphor of a `dial' that can be set to various values representing our commitments to various extra-empirical virtues that can decide between empirically-equivalent theories, and then writes (pp.73--75)
\begin{quote}
Wallace \ldots asserts that there is no underdetermination in
quantum mechanics, that there is only Everett. His argument is that Everett and only
Everett has been successfully applied to all of current physics. \ldots When the dial is set to include empirical reach or size of domain, there is no
underdetermination serious enough to cause alarm. \ldots For a quite different judgment, consider the position of Jean Bricmont [\citeyear{bricmontbook}]
when confronting quantum underdetermination. He argues that ``there is no existing
alternative to de Broglie-Bohm that reaches the level of clarity and explanatory power
of the latter''. \ldots we’re pretty close to philosophical bedrock at this
point. The Everettian and Bohmian described above aren’t merely disagreeing on the
correct dial setting, but they are disagreeing on the nature of the dial. Put somewhat
simplistically, the Everettian uses a dial that represents size of empirical domain
whereas the Bohmian uses a dial that represents explanatory virtues.
\end{quote}
Similarly, \citeN{eggnospeculation}, again discussing Bricmont and myself, writes (pp.3--4)
\begin{quote}
The problem of underdetermination underlies the notion of \emph{speculation} \ldots ontological content of a theory counts as speculative if it is subject to such underdetermination \ldots
On the one hand, Wallace [\citeyear{wallace-leedsrealism}] \ldots explicitly criticizes the claim that there is an underdetermination
here, because the Everett approach is the only one that takes the whole framework
of QM seriously, whereas its competitors (in particular, Bohmian mechanics and the
GRW theory) are almost exclusively concerned with a small subdomain of QM (namely,
non-relativistic particle mechanics). On the other hand, Jean Bricmont [\citeyear{bricmontbook}] argues
that none of the alternative versions of QM (Everett included) matches the Bohmian
approach in terms of clarity and explanatory power, which is therefore the only way to
really \emph{understand} QM. Ironically, then, by their very efforts to demonstrate the absence of
underdetermination, Wallace and Bricmont clearly show that one can with good reasons
hold on to one of at least two fundamentally incompatible versions of QM, which is to
say that any such choice is speculative in the sense employed here. [Emphasis in original.]
\end{quote}
So far as I can see, the common structure of these objections is:
\begin{enumerate}
\item There is underdetermination between the Everett interpretation and Bohmian mechanics.
\item That underdetermination could be resolved by extra-empirical virtues. But different authors disagree on the relative significance of the extra-empirical virtues.
\item For Wallace, the most important extra-empirical virtue is extendibility to the empirical domain of QFT.
\item For Bricmont, the most important extra-empirical virtues are clarity and explanatory power.\footnote{For the record, I don't concede the claim that Bohmian mechanics improves on the Everett interpretation in terms of explanatory power: depending on their attitude to the quantum state, the Bohmian either borrows the standard explanation of a quantum phenomenon like, \egc, superconductivity, and ignores the particles until the very end, or else black-boxes the quantum details entirely, since they cannot be spelled out in terms of primitive ontology. But I do not need this for my argument and so will not defend it in any detail.}
\item So Wallace and Bricmont's disagreement demonstrates the persistence of underdetermination.
\end{enumerate}
Stated that way, I hope the problem is obvious, but let me spell it out explicitly. \textbf{Empirical adequacy is not an extra-empirical virtue}. Extra-empirical virtues are, well, extra-empirical. They are introduced into the underdetermination debate to resolve ties between empirically equivalent theories. If two theories are not empirically equivalent, then there is no underdetermination, no ties to break, no need to consult the extra-empirical virtues. In Egg's terms, if `speculation' occurs when we commit to one theory over another even though there is no empirical distinction between them, there is no speculation involved in commitment to unitary quantum mechanics because it has no empirically equivalent rival once one allows for its predictive success in the QFT domain.

What \emph{seems} to be going on in this final strategy --- and, anecdotally, what seems to be going on in many philosophers' attitude to the supposed need to recover the predictions of QFT --- is that the underdetermination thesis is supposed to be already established once we observe that Bohm, dynamical collapse, and unitary quantum mechanics are empirically equivalent when restricted to the domain of NRQM.  It is, however, extremely unclear why this should matter. General relativity and Newtonian gravity are empirically equivalent when restricted to the non-relativistic domain, and Newtonian gravity is quite a lot simpler than general relativity. Come to that, quantum mechanics and \emph{classical} mechanics are empirically equivalent when restricted to the classical domain --- and classical mechanics certainly appears to  beat out quantum mechanics on simplicity grounds. But of course no-one is claiming that there is underdetermination between Newtonian gravity and general relativity, or between classical and quantum mechanics, with the wider empirical domain of one theory being weighed against the greater simplicity of the other! The discovery of new phenomena that one theory can predict but another cannot is the standard, go-to, \emph{empirical} move used by scientific realists to resolve apparent cases of underdetermination: it seems to apply here just as well as in other cases.

The best I can do in making sense of this idea is something like the following: 
\begin{quote}
Okay, we can't (yet) extend rivals to unitary QM to the QFT domain, but there is an enormous domain of physics to which they do apply, containing a vast number of novel phenomena, and comprising the great bulk of our evidence for QM. The ability of Bohmian mechanics, or dynamical collapse, to exactly reproduce the phenomena in this vast and rich domain is so remarkable as to give us strong reasons to think that in due course the theories can be extended to the QFT domain too, and so it's reasonable to discuss underdetermination on the hypothesis that this has been done.'
\end{quote}
Certainly, advocates of alternatives to unitary quantum mechanics are often keen to stress the breadth of what can be explained and predicted by their theories: for instance, Daumer~\emph{et al}~\citeyear{daumermessage}, in a polemical commentary on Anton Zeilinger's approach to quantum theory, refer to ``Bohmian mechanics \ldots a theory describing the deterministic evolution of particles that accounts for all of Zeilinger's examples and indeed all of the phenomena of non-relativistic quantum mechanics, from spectral lines to the two-slit experiment and random decay times''; similarly, the Stanford Encyclopedia entry on Bohmian mechanics (written by Sheldon Goldstein, one of the authors of Daumer~\emph{et al}) claims that Bohmian mechanics ``accounts for all of the phenomena governed by non-relativistic quantum mechanics, from spectral lines and scattering theory to superconductivity, the quantum Hall effect and quantum computing''. \cite{goldsteinsep}

But the scope of modificatory theories is narrower by far than this, as we will now see.

\section{Quantum electrodynamics and the scope of non-relativistic quantum mechanics}\label{scope-of-nrqm}

The term `non-relativistic quantum mechanics' is ambiguous. On one reading, it means simply the formal theory of non-relativistic quantum particle mechanics (NRQM), in which finitely many particles interact via electrostatic and magnetostatic potentials. But it can also mean the phenomenological regime in which energies are low and the exotica of special relativity and high-energy physics can be neglected. In this second sense of the term, non-relativistic quantum mechanics might be understood as excluding particle accelerators, or nuclear power stations, or stars, or cosmic rays, but as including pretty much all the ordinary, mundane, easily-accessible quantum phenomena of our slow-moving, low-energy world.

This anbiguity would be harmless if the scope of the two coincided --- if those `ordinary, mundane, easily-accessible quantum phenomena' in fact lie within the zone of predictive power of NRQM. But they do not: most of the most elementary applications of quantum mechanics lie outside the domain of phenomena describable by NRQM alone. I begin with the most elementary of all --- an empirical question so simple even a child could ask it.

Why is the sky blue? The answer\footnote{Again, I do not attempt to give original citations for well-established physics in this section. } is well understood: it is blue because air molecules preferentially scatter radiation towards the blue end of the visible spectrum. The relevant physics is \emph{Rayleigh scattering} (see, \egc, \citeNP[pp.462--471]{jackson}), appropriate when light is scattered off particles that are very small compared to the wavelength of light. Molecules are $\sim 10^{-10}$ meters across, and visible light has a wavelength of $~10^{-7}$ meters, so Rayleigh scattering applies, so air is blue, so the sky is blue. 

Neither Bohmian mechanics, nor the GRW theory, can reproduce this answer, for the obvious reason that it concerns \emph{light}, and light is not in the domain of NRQM. If you want an example of an experiment which resolves the supposed underdetermination between unitary QM and its rivals, you do not need particle accelerators or telescopes or equipment of any kind: you just need to look up.

One might respond\footnote{Thanks to John Norton for this observation.} that Rayleigh scattering is a classical phenomenon, and \emph{no} quantum theory needs to reproduce its results. I don't find this compelling. For one thing, all of these approaches to QM are committed to the idea that classical mechanics is an approximating or limiting case of quantum mechanics, so that the classical explanation must be underpinned by some quantum account. But more practically, classical accounts of Rayleigh scattering idealize molecules as perfectly reflecting spheres, which they are not: the classical model gives a very satisfactory qualitative account of why the sky is blue, but if you want to calculate exactly \emph{how} blue it is, you require a quantum electrodynamical treatment of Rayleigh scattering. It is not especially difficult to provide one in unitary quantum theory (cf \citeNP[pp.374--377]{loudonbook}) --- but it cannot be done in Bohmian mechanics or dynamical-collapse theory, until and unless those accounts can be generalized to quantum field theory.

The sky is not the only colored thing. We have a rich, and thoroughly quantum-mechanical, understanding of the reflectivity and color properties of solids and liquids (see \cite{nassaucolor} for an introduction). None of this is replicable in Bohmian mechanics or dynamical-collapse theories.

For a somewhat-related example, consider the spectral lines of atoms (singled out, recall, as an example of phenomena reproducible by Bohmian mechanics by Daumer~\emph{et al}~\citeyear{daumermessage} and \citeN{goldsteinsep}). Here the core \emph{calculation} --- the spectrum of the atomic Hamiltonian, for monoelectronic systems; the effective spectrum in a Hartree-Fock approximation, for multielectron systems --- is carried out in non-relativistic quantum mechanics\footnote{Mostly.  Some heavy atoms or for precision measurements require a relativistic treatment of the electrons, and the Lamb shift is a quantum-field-theoretic phenomenon.}. But spectral lines are defined by emission frequencies of light --- in elementary treatments we just apply $E=\hbar \omega$ (and pay a bit of attention to angular momentum conservation), but for a more careful treatment we need to couple the atom to the quantized electromagnetic field. Once again, this lies outside the scope of Bohmian mechanics or dynamical collapse theory.

Can we interpret spectral-line measurements just as measurements of energy levels (a non-relativistic phenomenon) and ignore the practical method we use to measure them (spectral-line frequency)? Not in Bohmian mechanics,  nor in dynamical-collapse theories, for the reasons developed in section \ref{what-is-required}: these theories are, and need to be, committed to modelling the actual measurement process, up to the level at which unitary quantum mechanics would describe macroscopic superpositions in position degrees of freedom. To review briefly: in either case we recover the Born rule only for probability distributions over positions (for dynamical-collapse theories, only over \emph{macroscopic} positions). Those do not arise in spectral-line measurements, where we are carrying out an energy measurement, not a position measurement. Those measurement results are transmitted into macroscopic position data by the physics of measurement --- but that physics lies outside the scope of non-relativistic quantum mechanics, and so outside the scope of either rival to unitary QM. As long as we confine ourselves to the physics of individual atoms, the Bohmian particles are epihenomenal and the dynamical-collapse mechanism is untriggered. 

The two-slit experiment provides a rather different example. Of course, it is usually done with photons, in which context it manifestly lies outside the scope of NRQM. But it can be done with massive particles, say electrons: indeed, \cite[pp.10-14]{maudlinqmbook} is careful to present the two-slit experiment in this way in his introduction to the philosophy of quantum mechanics. Thus performed, the core physics of the experiment indeed can be modelled within NRQM. But the measurement of the electrons' position at the end of the experiment cannot. Electrons are standardly\footnote{See, \egc, \cite{faruqihenderson}.} detected by one or other process that causes the electron to scatter a number of photons: in Maudlin's own presentation, for instance, a phosphor screen is used, and a flash of light marks the detection of each electron. Once again, this is well understood as a phenomenon in QED, but it cannot be modelled inside NRQM. (Note that in Maudlin's own subsequent discussion he is repeatedly explicit (e.g. pp.49-50, pp.68-69) that a satisfactory version of quantum theory absolutely has to be able to model the detection process in this way.)

Our three examples illustrate three ways in which apparently non-relativistic quantum measurements actually requires QFT. The first (illustrated by the color of the sky) is when the actual predicted result is not expressible in NRQM (normally it is some claim about frequencies or intensities of light). The second (illustrated by spectral lines) is when the quantity being measured can be described in NRQM but is not position, and where the process of measurement requires QFT to be modelled. The third (illustrated by the two-slit experiment) is when the quantity being measured is the position of non-relativistic particles but where the measurement process nonetheless uses QFT to magnify this to a macroscopic superposition.

For a quantum experiment to be modellable entirely within NRQM (and thus replicable within Bohmian mechanics or dynamical-collapse theory) not only the system being measured, but the apparatus doing the measurement, would have to be within the scope of NRQM. Such systems plausibly exist: the original Stern-Gerlach experiment\footnote{See, \egc, (Schmidt-B\"ocking \emph{et al}~\citeyearNP{sterngerlachreview}).}, for instance, splits a beam of silver atoms, deposits the atoms on a sheet, and waits until enough atoms have collected to be seen with the naked eye. That's probably modellable within NRQM from start to finish (though a sufficiently harsh critic might worry about the nature of the atom/sheet interaction and whether the slowing down of the silver atoms requires radiative emission). But experiments like this comprise only quite a small fraction of the experiments performed within `non-relativistic' quantum mechanics. Quantifying \emph{how} small a fraction is difficult, not least since the distinction as to which experiments do and do not require QED to model the measurement apparatus is not a scientifically-interesting one from the perspective of mainstream physics. But as a small illustration, let's consider the list of eight experiments used to explain quantum mechanics by \citeN{maudlinqmbook} (who is committed to the claim that all eight are replicable within Bohmian mechanics), the list of applications we saw previously from Daumer~\emph{et al}~\citeyear{daumermessage} and \citeN{goldsteinsep}, and all the experiments discussed in chapter 4 (`the early development of quantum mechanics') of \cite{Rae1992}. The first two lists are by advocates of QUT and so ought to offer relatively favorable ground for underdetermination; the third is from a very widely used undergraduate text.

To begin with Maudlin: his eight experiments (detection of single particles; one-slit and two-slit diffraction; two-slit diffraction with monitoring; the Stern-Gerlach experiment;  the Mach-Zender interferometer; the EPR experiment; Bell inequality violation) are all described using electrons and position measurements of those electrons, even when this is sharply at variance with how the experiments are actually performed (for instance,  EPR-type experiments are almost invariably carried out using photons; Mach-Zender interferometry with fermions is normally done with neutrons, not electron; the original Stern-Gerlach experiment uses silver atoms). I assume Maudlin does so to keep the experiments non-relativistic and as close as possible to Bell's idea of all measurements being measurements of position; still, none of these eight experiments are actually reproducible within Bohmian or dynamical-collapse theories, because they all use a phosphor screen to translate electron impacts into bursts of photons. Shifting to a more realistic description of the experiments mostly does not help (photon-based realizations are of course essentially QED phenomena; neutron detection generally\footnote{See, \egc, \cite{peurrungneutrons}.} involves induced radioactivity, which lies outside the scope of NRQM) although ironically a more realistic description of the Stern-Gerlach experiment probably does lie within NRQM.

Combining Daumer \emph{et al}'s list with Goldstein's gives us the following (I'll skip quantum computation as we don't really have any current experimental implementations of a quantum computer, so as yet they are not relevant to underdetermination):
\begin{enumerate}
\item Spectral lines: we have already seen that these cannot be reproduced within extant Bohmian or dynamical-collapse theories.
\item The two-slit experiment: likewise (whether we perform the experiment with photons, or electrons, or neutrons).
\item Random decay times: random decay is mostly applied to nuclei and other subatomic particles, which manifestly lie outside NRQM. The main non-relativistic example would be decay of excited atoms by emission of photons, but this is a QED phenomenon. I don't know of any experimentally-relevant example of a non-relativistic bound system decaying through emission of some non-relativistic particle.
\item Scattering theory: a really large fraction of experiments involving scattering theory, even in so-called `non-relativistic' physics, will be outside the scope of NRQM (and hence not reproducible by Bohmian or dynamical-collapse versions of NRQM). The exceptions will (a) involve collisions where radiative effects can be ignored, and (b) use particle-detection mechanisms that NRQM can treat. The application of scattering theory is so widespread that I am sure there are some examples, though I have not been able to come up with any myself.
\item Superconductivity: the calculational core of the BCS model of superconductivity can be performed entirely in NRQM. I'm less sure about the general framework of superconductivity: Weinberg'selegantly general derivation using effective-field-theory methods \cite{weinbergsuperconductivity}  relies on a quantized gauge field, for instance. Some but not all of the experimental signatures of superconductivity can be reproduced in NRQM: the iconic prediction of zero conductivity can; the photon's acquisition of mass
 elegantly general derivation cannot. (The dynamical-collapse mechanism, and the Bohmian particles, play no role in the explanation of BCS theory, but that's an extra-empirical vice and not directly relevant here.)
\item The quantum Hall effect: The core physics of the quantum Hall effect can be analyzed within NRQM; the ability of Bohmian mechanics and dynamical-collapse theory to reproduce any given quantum Hall experiment will therefore depend on the details of the measurement process, but I am happy to stipulate that at least some applications will be reproducible. (The celebrated use of the quantum Hall effect for precision measurements in QED\footnote{See, \egc, \cite{vonklitzing}, and references therein.} of course cannot.).
\end{enumerate}
Finally, here is Rae's list:
\begin{enumerate}
\item The photoelectric effect: involves photons, not within the scope of NRQM.
\item The Compton effect: likewise, involves photons.
\item Line spectra and atomic structure: we have already considered this case.
\item Electron diffraction in the Davisson-Germer experiment: actually might lie within the scope of NRQM since the detection mechanism\footnote{See, \egc, \cite{davissongermerreview}.} is crystallization on a nickel surface.
\item Neutron diffraction in modern experiments: neutron detectors cannot be modelled in NRQM.
\item Diffraction of buckminsterfullerene: the detection process~(Arndt~\emph{et al}~\citeyearNP{fullerene-diffraction}) involves photoionization, which cannot be modelled in NRQM.
\end{enumerate}

In all, there are a few partial, and probably one or two full, cases in our combined list of experiments where NRQM can adequately describe both the experiment and the measurement process, but they make up a fairly small minority. I hope the point is clear: even if (for no very good reason) we confine our attention to the regime of low-energy, tabletop physics, at most a rather small and fragmentary portion of the experimental evidence for quantum mechanics is modellable  entirely within non-relativistic quantum mechanics and hence reproducible within either Bohmian mechanics or current dynamical-collapse theories. Photonic physics is intimately connected to the physics of non-relativistic matter: the clean separation of NRQM and QED in physics education is done for pedagogical reasons (QED is much more difficult) and does not reflect any clear delineation of domains.

\section{Rival research programs?}\label{research-programs}

There is one more strategy available to the advocate of QUT: acknowledge that Bohmian mechanics and dynamical collapse theory are at present empirically inadequate, but reconstrue them --- and indeed unitary quantum mechanics --- as \emph{research programs}, uncompleted but continuing strategies to develop physical theories. \citeN[p.59]{callenderquarantine} advocates this fairly explicitly, construing each research program (Everett, Bohm, Collapse) in Lakatosian terms \cite{lakatosmethodology} with an unrevisable hard core but a revisable periphery that leads to a succession of new theories. \citeN[p.7]{eggsaatsi} hint at something similar when they suggest that ``the alternative research programs [to develop Bohmian and collapse versions of QFT] are advanced enough to count as genuine rivals to an Everettian account of QFT''.

I think this is probably the right way to think about the three alternatives, although (in keeping with the idea that we are considering underdetermination of \emph{theories}, not of interpretations) I would recharacterize the `Everett' research program as the `unitary quantum mechanics' research program, whose hard core is the unmodified quantum formalism, the universality of unitary dynamics, the commitment to modelling measurement physically rather than as some unmodelled external intervention, and the appeal to decoherence to understand the quantum/classical transition'. The Everett interpretation is then an interpretation of the theories within that research program --- probably the only viable interpretation in my view, but again, I'm not arguing for that here --- but it is a pure interpretation and does not modify the formalism. 

The first thing to note here is that if we are to understand the Everett / Bohm / Collapse controversy as a rivalry between research programs in this sense, the threat of underdetermination is largely defused. Rival research programs are common in science, and the rivalry can seldom be resolved by an unambiguous Popperian falsification, since apparent falsifications can normally be accommodated by modifying auxiliary assumptions while leaving the hard core of the program unchanged. Indeed, Lakatos introduced his framework of research programs, and the related idea that a research program can be `progressive' or `degenerating' depending on how it updates itself to deal with experimental anomalies, exactly to address this deficiency in Popper's original account. The dark-matter research program aims to explain the observed motions of stars and galaxies via a new form of non-baryonic matter; the modified-gravity research program aims to do so via changing Newtonian gravity; no deep problem of underdetermination arises if a given point in physics history there is not a conclusive case for one or another. Similarly, the big bang and steady-state hypotheses were incompatible cosmological theories and yet for some while the data did not decisively tell between them; that shows only that science is an ongoing process, not that there is any lasting problem for realism.

Nor does the mere existence of a research program tell us anything one way or another about whether the broader physics community thinks that program is empirically viable. The strong majority view in cosmology (cf \citeN{weatheralldarkmatter}) is that current evidence --- notably  the fluctuation structure of the cosmic microwave background and the observations of the Bullet Cluster --- decisively favors dark matter over modified gravity, but there still remains a small modified-gravity research program. The \emph{overwhelming} majority view in cosmology is, and has been for decades, that observation conclusively supports the big bang over the steady state hypothesis, but a trickle of papers in the steady-state research program were still appearing in peer-reviewed astrophysics journals into the 21st century (e.g.,~\citeNP{narlikarburbidgevishwakarma}).

And if we are to use Lakatos's philosophy to compare research programs in quantum mechanics, the results are not favorable to the Bohm and Collapse programs. There is something of a tendency to use `progressive' and `degenerating' as general-purpose terms of approbation or disdain, but Lakatos gives them a fairly precise meaning very tied to their empirical fruitfulness: a research program is progressive if its successive modifications tend to lead to novel confirmed predictions, degenerating otherwise. By that measure, the unitary quantum mechanics research program has been staggeringly progressive: its successive developments --- the Dirac and Klein-Gordon equations, QED, current algebra, electroweak symmetry breaking, chromodynamics, effective field theory, and many more --- have produced a wealth of novel predictions unmatched in breadth and accuracy in physics, and probably in science writ large. The Bohm and Collapse research programs have nothing even remotely comparable\footnote{Callender (p.73) suggests certain applications of Bohmian and collapse theories to quantum gravity, and notes that `[i]f Bohmian or Collapse answers to problems in new realms can't be reproduced by Everett then it's not clear who is more progressive'. He freely concedes that QFT `dwarfs these examples in importance' but I think the more important point is that these examples, unlike the predictions of QFT, \emph{are not empirical} --- they do not lead to novel confirmed predictions.} even if --- generously --- we allow already-predicted novel consequences of unitary quantum mechanics to count as `novel' if reproduced by a rival theory.

Callender, I think, would object. He writes that
\begin{quote}
the \emph{very features that allow the Everettian interpretation its easy
extension to new physics are precisely the same features that invite its problems and
whose solutions by other programs lead to their explanatory virtues.}\ldots The Everettian
program is a ‘minimalist’ one, little more than the quantum formalism itself coupled to
a new rule for reinterpreting our definite empirical outcomes. So it is little wonder that
it can be ‘successfully applied’ to new physics. To someone engaged in a competing
research program, trumpeting Everett's easy application to new physics sounds like
a thief bragging about how little they had to work for their reward. [\cite[pp.74-5]{callenderquarantine}, emphasis in original.]
\end{quote}

Callender's reference to a 'thief' is, I assume, an allusion to Russell's famous  observation \cite[p.71]{russell1919} that
\begin{quote}
The method of `postulating' what we want has many advantages; they are the same as the advantages of theft over honest toil.
\end{quote}
But from the Everettian perspective, the `honest toil' has already been done, over generations and with spectacular results, by the thousands of physicists who have developed the unitary quantum mechanics research program, and the desire for an \emph{interpretation} of the unitary quantum formalism and not a \emph{modification} comes from a mixture of respect for this toil and belief that it would be remarkable --- a miracle, indeed --- if it had produced the fruits it has without being basically correct. The physicists who have tried to actively develop Bohmian and dynamical-collapse versions of QFT start from the viewpoint that the unmodified theory \emph{cannot} be correct, and so the work of developing post-1930 physics must be done anew, and if they have little to show for it so far, still their toil is no less honest than their rivals'. But as for the strategy of simply postulating that Bohmian or dynamical-collapse versions of QFT can be developed to the point of being empirically equivalent to unitary quantum mechanics, and then arguing for underdetermination on the basis of that postulate\ldots well, Russell said it more clearly than I can.

\section{Conclusion: underdetermination and miracles}\label{conclusion}

Unitary quantum mechanics stands unrivaled as our empirically most successful theory of the quantum domain. No extant version of Bohmian mechanics, and no extant dynamical collapse theory, succeeds in reproducing more than a small fragment of its predictions, even if for some reason we choose to restrict ourselves to non-relativistic energies.  If the Everett interpretation is the only viable interpretation of unitary quantum mechanics, then it is the only currently-available way to understand the quantum world; if instead there are other viable interpretations of the unmodified unitary formalism, then theere may be underdetermination between those different interpretations; but there is no underdetermination of \emph{theory} by evidence in quantum mechanics. The quantum underdetermination thesis is false.

What if, someday, a new hidden-variable or dynamical-collapse theory is developed, one that is clearly incompatible with unitary quantum mechanics and yet reproduces all its predictions, even in the domain of quantum field theory? Then we will have underdetermination of theory by evidence --- just as we would if someday we have a new proposal to explain cosmological redshift and the rest of the evidence for the big bang in a steady-state framework. But that day has not come, and I see no reason to expect it to come, for the same reason in either case: our current best theory is so successful, predicts so much novel confirmed empirical data, that it would be miraculous if it was not at least approximately the right story about how the world is structured and organized in its domain of application; if an incompatible alternative existed which made the exact same predictions, then it would be likewise miraculous if that alternative was not at least approximately the right story; so if underdetermination occurs then at least one story or other is miraculously successful despite its falsity; so if you don't believe in miracles, you shouldn't believe in underdetermination. \emph{Actual} underdetermination would give us strong grounds for scepticism about the case for scientific realism, but absent actual underdetermination, that same case gives us good reason for scepticism about merely possible underdetermination.

Advocates of Bohmian mechanics, and of dynamical collapse, will reject this argument, for reasons that are entirely proper given their commitments. In general they reject the claim that the Everett interpretation offers an adequate interpretation of unitary quantum mechanics; indeed, they deny the existence of \emph{any} adequate interpretation of unitary quantum mechanics. They accept Bell's dichotomy: ``either the wavefunction, as given by the Schr\"odinger equation, is not everything, or it is not right'' \cite[p.201]{bellquantumjumps}. And so from their perspective, the task of constructing an empirically adequate alternative to unitary quantum mechanics is unavoidable, however daunting it may seem.

If one accepts this perspective, the contributions of de Broglie, and Bohm, and GRW, and Pearle, and Bell, are of inestimable value \emph{even though} they have not provided us with an empirically viable alternative to unitary quantum mechanics. In reproducing the general \emph{form} of the two slit experiment, the Mach-Zender interferometer, the Stern-Gerlach experiment, and so forth, they both provide proof of concept that quantum-type phenomena can in principle be explained in a way that sidesteps the (supposed) incoherences of unitary quantum mechanics, and act as a starting point from which we might hope to construct more realistic, empirically adequate, theories. At least in the case of dynamical-collapse theories (and perhaps also for the de Broglie-Bohm theory --- cf \cite{valentiniwestman}) they also provide a possible route to the Holy Grail of modificatory solutions to the measurement problem: empirical evidence against unitary quantum mechanics.

The concrete task of developing empirically adequate versions of either approach --- or any other empirically adequate alternative to unitary quantum mechanics --- will seem either compelling or quixotic, depending on one's optimism or pessimism about the prospects of overcoming the difficulties discussed in section~\ref{adequacy-in-qft} for extending either to QFT, and more importantly, on whether one thinks the measurement problem is in any case dissolvable via an Everettian (or other) interpretation of the unmodified quantum formalism. All sides ought to agree that the task is both profoundly demanding and scientifically honorable. It is ill served by complacency about how much has already been achieved. The philosophy of science is likewise ill served by underdetermination theses built not on actual empirical equivalence between theories we currently have, but on imagined empirical equivalence between our current best theory and our dreams of the theories we aspire one day to create.

\section*{Acknowledgements}

I am grateful for feedback on earlier versions of this work from audiences in Oxford, Pittsburgh, and at two virtual conferences (`The Quantum Limits of Knowledge' and `On the Shoulders of Everett').


\end{document}